# DHLP 1&2: Giraph based distributed label propagation algorithms on heterogeneous drug-related networks

Erfan Farhangi Maleki
[1]Department of Electrical and Computer Engineering,
Isfahan University of Technology,
Isfahan 84156-83111, Iran
[2]Department of Computer and Information Sciences,
University of Delaware,
Newark, DE, 19711, USA
erfanf@udel.edu

Nasser Ghadiri *
Department of Electrical and Computer Engineering,
Isfahan University of Technology,
Isfahan 84156-83111, Iran
nghadiri@cc.iut.ac.ir

Maryam Lotfi Shahreza
Department of Computer Engineering,
University of Shahreza,,
Shahreza 86149-56841, Iran
maryam.lotfi@gmail.com

Zeinab Maleki
Department of Electrical and Computer Engineering,
Isfahan University of Technology,
Isfahan 84156-83111, Iran
zmaleki@cc.iut.ac.ir

**Abstract:**

*Background and Objective:* Heterogeneous complex networks are large graphs consisting of different types of nodes and edges. The knowledge extraction from these networks is complicated. Moreover, the scale of these networks is steadily increasing. Thus, scalable methods are required.

*Methods:* In this paper, two distributed label propagation algorithms for heterogeneous networks, namely DHLP-1 and DHLP-2 have been introduced. Biological networks are one type of the heterogeneous complex networks. As a case study, we have measured the efficiency of our proposed DHLP-1 and DHLP-2 algorithms on a biological network consisting of drugs, diseases, and targets. The subject we have studied in this network is drug repositioning but our algorithms can be used as general methods for heterogeneous networks other than the biological network.

*Results:* We compared the proposed algorithms with similar non-distributed versions of them namely MINProp and Heter-LP. The experiments revealed the good performance of the algorithms in terms of running time and accuracy.

# 1 Introduction

Complex networks are graphs with non-trivial and complicated structural features that do not occur in simple networks such as lattices or random graphs. Modeling different processes with complex networks has recently attracted the research community (Silva & Zhao, 2016).

Most real-world networks such as social networks and biological networks are modeled as heterogeneous networks, which consist of different types of nodes and edges and make the knowledge discovery from such networks complicated and time-consuming. In comparison with homogeneous networks, heterogeneous networks contain richer structural and semantic information. Therefore, gaining knowledge and mining such networks requires specific algorithms with features different from the algorithms that run on homogeneous networks. On the other hand, their growth rate is much higher than that of homogeneous networks. Therefore, with the advent of such networks and considering the heaviness of the required processes, some algorithms and platforms are required to provide better performance and scalability in the face of such structures.

There are different approaches to discovering knowledge in heterogeneous networks, including semi-supervised learning. Label propagation is among the well-known and successful methods in this domain (Silva & Zhao, 2016). Its strength is in utilizing both local and global features of the network for semi-supervised learning (Zhou, Bousquet, Lal, Weston, & Schölkopf, 2004). In different approaches of label propagation, specific labels are assigned to individual nodes of the network, and the label information is then repeatedly propagated to the adjacent vertices. The propagation process is finally converged toward minimizing the objective function (Shahreza, Ghadiri, Mousavi, Varshosaz, & Green, 2017).

Bulk Synchronous Parallel (BSP) (Valiant, 1990), which is a parallel and vertex-centric programming model, has been used by Malewics in Pregel system. Google has introduced Pregel and implemented in C/C ++ language for large-scale processing of graphs (Malewicz et al., 2010). The computations in Pregel are carried out by a sequence of super-steps. In each super-step, every node that is involved in calculations 1) receives the sent values of adjacent nodes from the previous super-step, 2) updates its values and state, and 3) sends its updated values to its neighboring nodes, which will be available in the next super-step. The Apache Giraph framework is an iterative system of graph processing inspired by Pregel. Giraph is an open-source platform that executes on the Hadoop distributed infrastructure to conduct the computations on billions of edges and thousands of machines (Martella & Shaposhnik). Giraph has developed the initial model of Pregel with enhanced features such as out-of-core computation, master computation, shared aggregators, and combiners (Ching, Edunov, Kabiljo, Logothetis, & Muthukrishnan, 2015).

In the present paper, due to the iterative nature of label propagation algorithms, we selected Apache Giraph as a distributed graph processing platform that makes use of vertex-centric programming model and is a good fit for iterative and scalable algorithms.

The MapReduce programming model and the Hadoop distributed processing framework are designed mainly for analyzing unstructured and tabular data (Farhangi, Ghadiri, Asadi, Nikbakht, & Pitre, 2017; Maleki, Azadani, & Ghadiri, 2016). However, they are not suitable for graph processing due to iterative nature of graph algorithms (Martella & Shaposhnik). Moreover, the results of experiments have revealed that iterative graph processing with the BSP significantly outperforms MapReduce especially for algorithms with many iterations and sparse communication (Kajdanowicz, Kazienko, & Indyk, 2014).

In addition to Giraph, other applications such as Haloop, Twister, Graphlab, Graphx, and Grace have been introduced to process iterative graph algorithms (Ching et al., 2015; Martella & Shaposhnik).

Giraph has one or more than one of the following advantages over each of the abovementioned applications:

1) Finding the cause of a bug is faster and easier in Giraph.
2) Giraph is more memory-efficient than other methods, and the problem of the out-of-memory process occurs less frequently. Even if this happens in Giraph, it can conduct computations thanks to the out-of-core feature it has.

3) Unlike some applications such as Pregel, Giraph is open-source.
4) It performs better than these applications for higher volumes of data
5) In comparison with some other platforms, it has less overhead in using the network.

Regarding the fact that many business and research institutions use Hadoop, making use of other systems requires the creation of a separate service to work with the graph, while Giraph is implemented on Hadoop, and this is not needed. Furthermore, Giraph has been written in Java, whereas some of the other applications like Graphlab is written in C/C++ and as a result, there is less compatibility in them.

We will evaluate the proposed algorithms and platform in practical usage in the bioinformatics domain for "drug repositioning". Discovering and developing new drugs through clinical experiments is a time-consuming and costly process. Most drugs fail during discovery and development while using available drugs to cure diseases other than those they are developed for, involves lower risk, cost, resources and a waste of time. This process is called "drug repositioning" scientifically. Actually, the research objective of this paper is to propose scalable label propagation algorithms to reduce the time of knowledge discovery in heterogeneous biological networks with the application in drug-repositioning. To the best of our knowledge, we are the only contribution that presents a scalable implementation of label propagation algorithms for heterogeneous drug-related networks.

The evaluation process mentioned above consists of two steps: 1) Creating a heterogeneous network with Giraph input format: our initial input includes drug similarity network, disease similarity network, target similarity network, known drug-disease interactions, known drug-target interactions, and known disease-target interactions. In other words, we have three similarity matrices and three interactions matrices that must be first integrated into the Giraph input format for running our algorithms on them. 2) Predicting the presence of potential interactions: in this step, two distributed heterogeneous label propagation algorithms called Distributed Heterogeneous Label Propagation 1 and 2 (DHLP-1 and DHLP-2) are developed so as to discover potential interactions of drug-target, drug-disease, and disease-target at the right time and in a scalable way. The runtime of the algorithms is measured for this purpose on various datasets with different sizes and compared with the non-distributed versions of the algorithms.

The first output of the proposed algorithms is interactions matrices of drug-disease, drug-target, and disease- target. The second output includes new similarity matrices for drugs, disease, and targets. The final output is sorted lists of candidates for drug repositioning. Unlike the existing methods, the algorithms proposed in the paper can identify new drugs interactions (drugs without the corresponding target) as well as new targets (targets without the relevant drug). In this paper, some experiments have been designed to show this ability. Moreover, there is no need for negative training sampling in such algorithms.

The experiments have been designed based on 10-fold cross-validation to evaluate the accuracy of the proposed algorithms. The analysis is carried out according to widely-used performance metrics of AUC, AUPR, and best accuracy.

The rest of this paper is organized as follows: In Section 2, we review the related work. The complete description of our proposed methods is introduced in Section 3. The time complexity of the algorithms is investigated in Section 4. In Section 5, the regularization framework and proof of the convergence of the algorithms are presented. The performance evaluation of the algorithms is provided in Section 6, and Section 7 gives a summary of the research.

## 2 Background

Research on heterogeneous networks has significantly expanded over the past years. The existing methods for extracting knowledge from the networks could be categorized into the domains of measuring the similarity, clustering, classification, predicting the presence of an edge, ranking, and recommending (Shi, Li, Zhang, Sun, & Philip, 2016).

There are various approaches to extract knowledge from complex networks. For example, a linguistic knowledge extraction with complex networks is proposed using clustering as a part of the process (Stanisz, Kwapień, & Drożdż, 2019). besides, one of the existing semi-supervised algorithms for acquiring knowledge from the complex networks such as heterogeneous networks is the Label Propagation (LP) algorithm that is closely related to the Random Walk (RW) algorithm (Grady, 2006). However, there are two differences: 1) LP fixes the labeled points, and 2) the LP's response is an equilibrium state while RW's output is dynamic. The label propagation algorithm is mainly used for community detection (Gregory, 2010; Tian & Kuang, 2012; Xie & Szymanski, 2013) but can also be used for link prediction (Liu, Xu, Xu, & Xin, 2016) and text classification

(Rafael Geraldeli Rossi, de Andrade Lopes, & Rezende, 2016; Rafael G Rossi, Lopes, & Rezende, 2014). In label propagation for *homogeneous* networks, the labels are propagated only in one network consisting of the nodes and edges with the same type such as the work carried on in (Zhou, Bousquet, Lal, Weston, & Schölkopf, 2004). A current challenge is how to propagate the information in the heterogeneous networks consisting of several subnetworks. heterogeneous networks contain richer structural and semantic information. Each subnetwork has its own clustering structure that has to be analyzed independently. In label propagation for *heterogeneous* networks, the label is propagated in a network consisting of the nodes and edges with different types. Therefore, it has to be adaptable to these kind of networks and support the inherent nature of the heterogeneous networks while label propagation for homogeneous networks is not able to fulfill. Discovering the disease-gene interaction (Hwang & Kuang, 2010), detecting drug-target interaction (X. Chen, Liu, & Yan, 2012; Yan, Zhang, & Zhang, 2016), and drug repositioning (Shahreza, Ghadiri, Mousavi, Varshosaz, & Green, 2017) are among the research works on this issue and have been appropriately used in biological subjects. In (Hwang & Kuang, 2010), an algorithm called MINProp and a regularization framework is introduced for label propagation over the subnetworks of a heterogeneous network. MINProp sequentially conducts label propagation in each subnetwork using the current label information that has received from other subnetworks. The algorithm runs until convergence, and the global optimum of the target function is achieved. In (Yan et al., 2016), an algorithm named LPMIHN is employed to discover the possible relations of drug-target using the heterogeneous network. In this algorithm, label propagation is done in each heterogeneous subnetwork separately, and interactions of heterogeneous subnetworks are used only as extra information to form similarity matrices. Also, the presence of a large number of repeating loops limits its use for large datasets.

In (X. Chen et al., 2012), three subnetworks of protein similarity network, drug-similarity network, and known drug-target interaction network are first integrated and form the heterogeneous network. Next, Random Walk is utilized to discover new drug-target interactions.

Heter-LP (Shahreza, Ghadiri, Mousavi, Varshosaz, & Green, 2017) is another label propagation algorithm for heterogeneous networks. It was presented as a general algorithm and evaluated for drug repositioning problem by applying on an integrated network composed of six subnetworks (drug-drug, drug-disease, drug-target, disease-target, disease-disease, target-target). Different analysis performed in (Shahreza, Ghadiri, Mousavi, Varshosaz, & Green, 2017) shows improved accuracy in predicting new drug-disease, drug-target, and disease-target interactions.

Because heterogeneous networks are naturally large-scale and performing algorithms like label propagation in such networks will require many iterative calculations, making use of particular distributed graph processing platforms for label propagation can provide much higher efficiency. The concept of distributed label propagation for detecting communities in the homogeneous networks is suggested in a recent work (Bhat, 2012). Another proposed methodology is to parallelize the label propagation algorithm and the proposed similarity measure using BSP programming model to perform community detection and link prediction in large-scale homogeneous networks (Mohan, Venkatesan, & Pramod, 2017). In the case of heterogeneous networks, little attention has been paid to the running time challenge. To give an example, distributed memory parallel algorithms for switching edges in massive heterogeneous networks is proposed by (Bhuiyan, Khan, Chen, & Marathe, 2017).

The existing computational approaches for research on drug repositioning include 1) using the methods that are based on machine learning (Gottlieb, Stein, Ruppin, & Sharan, 2011; Menden et al., 2013; Napolitano et al., 2013; Zhang, Wang, & Hu, 2014), 2) using text mining and semantic inference methods (Andronis, Sharma, Virvilis, Deftereos, & Persidis, 2011; B. Chen, Ding, & Wild, 2012; Tari & Patel, 2014; Zhu, Tao, Shen, & Chute, 2014), and 3) using network analysis (Alaimo, Pulvirenti, Giugno, & Ferro, 2013; H. Chen & Zhang, 2013; Hwang & Kuang, 2010; Li & Lu, 2012; Wang, Yang, & Li, 2013; Xia, Wu, Zhou, & Wong, 2010; Yan et al., 2016). Our proposed methods are also based on a heterogeneous network analysis approach and the distributed versions of MINProp and Heter-LP algorithms. Our proposed methods DHLP-1 and DHLP-2 have the following advantages:

1) The speed and scalability in processing have not been touched on in any of the network analysis methods. Hence, their processing takes a long time. Our proposed methods have eliminated this shortage and show high speed and scalability in experiments rather than the non-distributed versions of them.
2) Our proposed methods have maintained the advantages of their non-distributed versions such as the ability to predict interactions with new concepts in the network, the ability to predict eliminated interactions, no need for negative samples and so on (Shahreza, Ghadiri, Mousavi, Varshosaz, & Green, 2017).

3) Our proposed methods show even better prediction accuracy compared their non-distributed versions (MINProp and Heter-LP). This results are described in Section 6.2.1,

## 3 Methods

As discussed in Section 1 and Section 2, we introduce two distributed label propagation algorithms for heterogeneous networks named DHLP-1 and DHLP-2. We use drug repositioning as the illustrative case study. Subsection 3.1 presents the formal notations and setting used in the problem. Subsection 3.2 covers data preparation. In Subsection 3.3 explanations about the framework used in the problem are provided. Subsection 3.4 explains the label propagation algorithms DHLP-1 and DHLP-2 and also Pseudo-Code of them are presented.

### 3.1 Notations and Setting

As can be seen in Figure 1, the heterogeneous network under investigation consists of three kinds of nodes: drugs, diseases, and targets. The edges between the nodes inside a homogeneous subnetwork show their similarity, and the edges between the nodes of two homogeneous subnetworks show their association. Therefore, there are three similarity-type edges including drug similarity network, disease similarity network, and target similarity network, and three association-type edges, i.e., known drug-disease interactions, known drug-target interactions, and known disease-target interactions.

The above-mentioned heterogeneous graph $G = (V, E)$ consists of three homogeneous subnetworks and three heterogeneous subnetworks. We show the homogeneous subnetwork as $Gi = (Vi.Ei)$ where $i = 1, 2, 3$ for drugs, diseases, and targets. The heterogeneous subnetworks are shown as $G_{i.j} = (V_i \cup V_j. E_{i.j})$ where $i, j = 1,2,3$ and $i \neq j$. $Ei$ indicates all edges that exist between the nodes of the homogeneous subnetwork, i.e. $Vi$. $Ei.j$ demonstrates all edges that exist between $Vi$ and $Vj$ nodes. Thus, for the graph $G$, we have following representation:

$$V = (V_1.V_2.V_3) . \quad E = (E_1.E_2.E_3.E_{1.2}.E_{1.3}.E_{2.3})$$

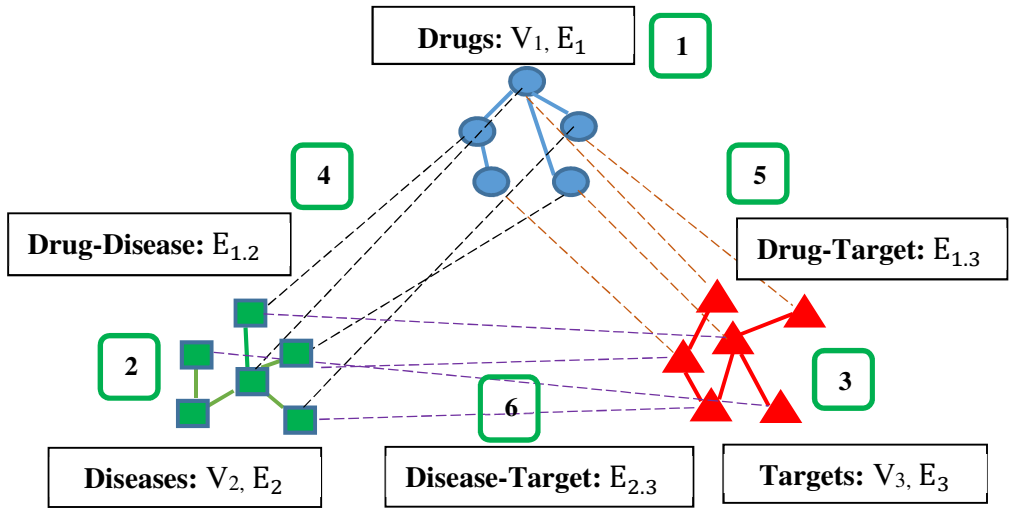

*Figure 1. The heterogeneous network model*

In this paper the inputs of the homogeneous subnetworks is represented by $P_i$ that is a proximity matrix with $n_i \times n_i$ dimensions where $n_i$ is equal to $|V_i|$, and $P_i(k.k') \geq 0$ is indicative of the similarity degree between entities k and k'. Moreover, the inputs of heterogeneous subnetworks is represented by $R_{i.j}$ that is a relation matrix with $n_i \times n_j$ dimensions where $n_i$ is equal to $|V_i|$ and $n_j$ is equal to $|V_j|$, and $R_{i.j}(k.k') \in \{0.1\}$ is indicative of the relation between entity $k$ from $i$th subnetwork and entity $k'$ from $j$th subnetwork. All $P_i$ and $R_{i.j}$ matrices must be normalized for the convergence of algorithms (Shahreza, Ghadiri, Mousavi, Varshosaz, & Green, 2017). Eventually, the normalized matrices are named $S_i$ and $S_{i.j}$.

## 3.2 Primary Data Set

In this section, we discuss the primary data set used to conduct the required processes. As said before, the heterogeneous network here consists of three concepts of drug, disease, and target. There are six matrices, three of which are similarity matrices, and the other three ones are binary association matrices. Similarity matrices indicate internal relations between the same type of entities, and association matrices show interrelations between entities of two different concepts. These datasets are gathered by the integration of gold standard and independent datasets. The gold standard datasets have been collected by (Yamanishi, Araki, Gutteridge, Honda, & Kanehisa, 2008) and include drug-target interactions, drug similarity, and target protein similarity. Four groups of proteins (Enzyme, GPCR, Ion Channel, and Nuclear Recep-tor) have categorized them. The primary resources of these data are KEGG, BRITE, BRENDA, SuperTarget, and DrugBank. The similarity of drugs is computed by SIMCOMP on chemical substructures, and the similarity of targets is computed by a normalized version of the Smith-Waterman score. However, due to the lack of information about diseases and their interactions with drugs and targets, more data have been added to them by suggesting the disease relationships with each of the four groups (Shahreza, Ghadiri, Mousavi, Varshosaz, & Green, 2017). This will constitute our primary data set.

In the next phase, preprocessing is required and some modifications should be made so that the data can be processed on Apache Giraph in a distributed way by our proposed algorithms. In step A of Figure 2, the required preprocessing is shown that will be described below, and in Section 3.3, it will be discussed in detail.

### 3.3 Workflow Description

In our heterogeneous network, three matrices exist for each concept. For example, the matrices related to the drug are: 1) the drug similarity matrix 2) the drug-disease binary association matrix and 3) the drug-target binary association matrix. The number of entities of each concept in different matrices is not the same in the primary data set. For example, there may exist some drugs in the drug-disease binary association matrix that do not exist in the drug similarity matrix or in the drug-target binary association matrix. Therefore, the number of entities of each concept in each related matrix should be the same in our proposed methods. For this purpose, data dimension homogenization is done for each concept in all three related matrices. The number of entities of each concept in each related matrix is equal to the number of distinct entities of each concept in the original matrices.

In the next step, the names of drugs, diseases, and targets are removed; only the values of interactions remain. There is no need to keep the name of every entity in the Giraph data format, and different entities are distinguished based on their Ids.

In other words, the input to Giraph is a graph, and every graph node will have a unique ID. Thus, an algorithm is required to allocate IDs to the entities of each type of concept. So, the IDs of drugs, diseases, and targets will be $3x + 1, 3x + 2$, and $3x + 3$, respectively where $x \in \{0.1.2.3. ... \}$. Hence, if a message from one of the adjacent nodes of one entity appears in running label propagation algorithms, that entity will be able to consider the sender's ID and determine to which concept this message belongs. After an ID is assigned to each entity, six matrices must be integrated into the form of an input file based on our Giraph input file format to prepare the file for processing.

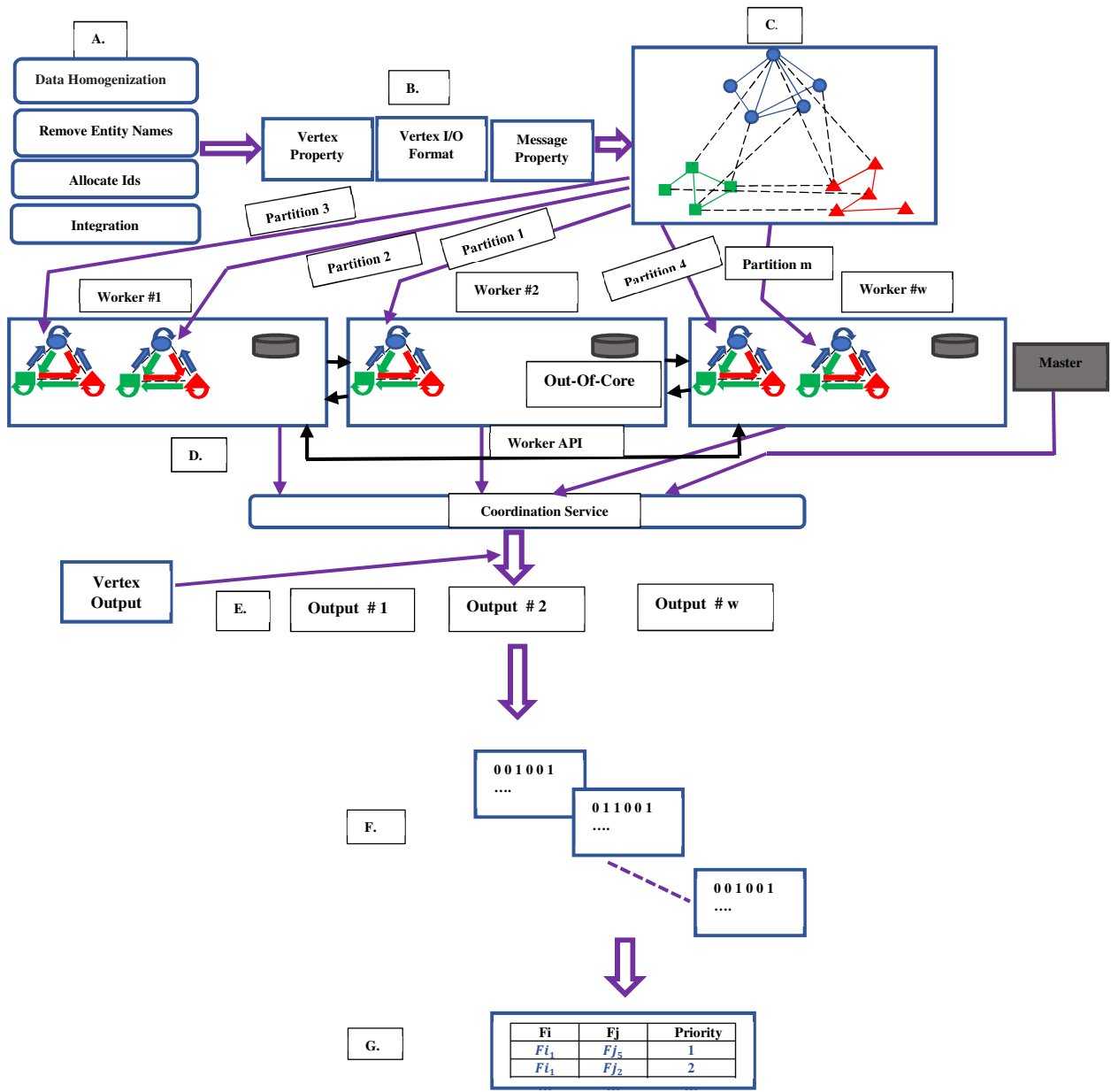

*Figure 2. The overall process workflow of DHLP-1 and DHLP-2 algorithms. A. Preprocessing is done on data B. Giraph-based structures are defined  C. The heterogeneous network is constructed and then given to Giraph as an input.  D. Distributed Label Propagation algorithms are performed on workers. E. The first output is generated. F. The second output is generated G. The sorted lists of predicted interactions are generated.*

Giraph allows us to define and implement our network data structures based on our processes. The four structures specified in this step include (step B in Figure 2):

1) Vertex Properties: in this stage, the vertex properties (such as local variables and arrays for storing information of the adjacent nodes) are determined.
2) Vertex Input Format: it is possible for us to determine Giraph input based on the initial inputs as well as our algorithm.
3) Vertex output Format: in this stage, the outputs resulting from running the algorithm are determined.
4) Message Properties: processing in Giraph is based on message-passing. In this stage, we determine what information the messages exchanged between the adjacent nodes in each super-step contain. The nodes receive the information, update their values and afterward, inform their adjacent nodes about their updated values through messages.

According to the Vertex Input Format, algorithm' data input is constructed in step C of Figure 2, which is a heterogeneous network consisting of three concepts. The data input in step D of Figure 2 is given to Giraph and is divided into different partitions each of which is a subset of the graph vertices and edges. Each partition is assigned to a worker to run the algorithm. In other words, one or more partitions are assigned to every worker. Dividing the graph into different partitions is considered a significant improvement in the runtime of the algorithm since they can be run on a worker simultaneously and in a parallel way. An ability called multi-threading has been embedded in Giraph based on which a user can maximize the calculations effectiveness and increase the number of the partitions that can be processed simultaneously by increasing the number of threads for each worker. As a result, the speed of algorithm execution also increases. However, if the number of the threads increase too much, the Giraph overload will be increased, and the application will become slow. Thus, there is an optimal value for increasing the number of threads depending the properties of the algorithm and hardware features. The effect of multi-threading on the speed of algorithms is shown in Figure 3. Each worker has a set of network APIs that allows remote workers to manipulate the data of their specific partitions. The workers are transferred from one super-step to the other by an active master. In each super-step, the workers search in all of their partitions, and they run the `Compute()` method for all of the vertices of such partitions. In addition to an active master, we have one or more standby masters that will replace the active master if it fails. The tasks of the masters include assigning the partitions to active workers, monitoring the workers statistical status, and transferring the workers from one super-step to the other. Coordination services are not involved in running graph processes. Instead, they provide distributed configuration, synchronization, and naming registry services.

After running the algorithm, the output is assigned to the central node based on the predetermined format (Step E of Figure 2). Furthermore, the output is divided between all of the workers, and every division contains a portion of the whole output. The output of each node includes the ID of the node and the final interactions values of its adjacent nodes. In the output files, the results are not placed using the order of ID numbers; the node that finishes its task sooner does not wait for the others and starts writing in the output file sooner than others. IDs in the form of *3x+1*, *3x+2*, and *3x+3* represent the drugs, diseases, and targets, respectively.

Moreover, the *x* value suggests how many entities of that concept precede them. In the next step (step F in Figure 2), drug-drug, drug-disease, drug-target, disease-disease, disease-target, and target-target matrices are generated. In the final step (step G in Figure 2), the interactions of each entity are sorted so that the interactions with the most similarities can be determined. As an example, for the drug-target matrix, the targets are determined for each drug according to their similarity degree. The new interactions will be recognizable as well.

### 3.4 The proposed label propagation algorithms

We propose two distributed heterogeneous label propagation algorithms to predict different types of potential interactions in the network efficiently and accurately. The heterogeneous network here consists of three concepts, namely drug, disease, and target and includes subnetworks of such concepts and the edges between them. In the naïve label propagation algorithm, some labels of nodes are known, and the aim is to estimate the label of other unlabeled nodes (Zhou, Bousquet, Lal, Weston, & Schölkopf, 2004). In each round, only the label of one node is specified, and the labels of other nodes will be eventually obtained after running the algorithm.

Some of the graph-based label propagation algorithms propagate the label only on one homogeneous network, so they are not suitable for heterogeneous networks because they are not able to capture the inherent nature of the heterogeneous networks due to the fact that heterogeneous networks contain richer structural and semantic information than homogeneous networks. The existing label propagation algorithms have focused on problem-solving and accuracy, and to the best of our knowledge, none of them considered the performance and scalability. Thus, running such algorithms is too time-consuming. In our algorithms to resolve those deficiencies, they have been proposed for heterogeneous networks and also distributed computing has been applied to speed up the algorithms.

Algorithms will be implemented in Giraph using vertex-centric programming. First, the `Compute()` method has to be implemented. In each super-step, Giraph calls this method for all active vertices and delivers the messages sent from the previous super-step to that vertex. The input of that method is the vertex and messages. The vertex determines the node and messages are a list of messages from previous super-step which are delivered to that node

The algorithm that is written in a vertex-centric language and is supposed to be run in a distributed platform must have the following features:

1) Vertices independent decision-making: each vertex contains a series of variables and local parameters. Decision-making of each vertex such as being halted, `voteTohalt()`, or sending messages to other vertices is only carried out based on local information.
2) Initializing vertex values: vertex values have to be initialized correctly. The path that calculations take depends on the structure of the graph as well as the initialization of the vertex values.
3) Halt conditions: each vertex makes the decision independently. Therefore, the halt conditions defined must be consistent and easily understandable. Moreover, a decision must be made on the effect of collaboration with other vertices.

The symbols used in this article are presented in Table 1, as well as explanations and pointing to the functions that use them.

*Table 1. Symbols used in this article and their descriptions*

| Symbol | Compute() | DHLP-1() | DHLP-2() | Description |
|---|---|---|---|---|
| `getSuper-step()` | ✓ | | | To get the current super-step number |
| `vertex.setValue()` | ✓ | ✓ | ✓ | To update and initialize vertices values |
| `vertex.getValue.gety()` | ✓ | ✓ | | To get the y value of the vertex |
| `PropagateMessage()` | ✓ | ✓ | ✓ | To propagate the vertex information to one-hop adjacent vertices |
| `vertex.voteToHalt()` | ✓ | | | To halt the vertex until it is activated again by received messages in next super-steps |

| Name | | | | Description |
|---|---|---|---|---|
| `vertex.getValue().getCurrentf()` | | ✓ | | To get the current value of f_t in DHLP-1 algorithm |
| `vertex.getValue().IsEnd()` | | ✓ | ✓ | To specify whether the vertex operation ends and the result is converged |
| `Flag` | ✓ | ✓ | ✓ | A flag which its zero value means the DHLP-1 and DHLP-2 algorithms can be executed. Its value is determined based on the execution of the early_checking() function |
| `vertex.getEdges()` | | ✓ | ✓ | To get the adjacent vertices |
| `vertex.getId()` | | ✓ | ✓ | To get the Id of a vertex |
| `edge.getTargetVertexId()` | | ✓ | ✓ | To get the Id of the target vertex for an edge |
| `edge.getValue()` | | ✓ | ✓ | To specify the value of an edge between two vertices |
| `Vertex.getValue().neighbours_f()` | | ✓ | ✓ | Each vertex knows the label values of the adjacent vertices through message propagation and then stores the values in its local memory. This function is used to get an array in which the label values of the adjacent vertices are returned. |
| `LPVertexValue` | ✓ | ✓ | ✓ | A data structure which specifies the features that each vertex must have in its memory that is determined based on the requirements of the program. Here includes feaures such as current lable of the vertex, lable in previous iteration, lables of neighbours and so on |
| `vertex.getValue().gety_prim()` | | ✓ | ✓ | To get the y' value |
| `F` | | | ✓ | To specify the label value of each vertex in the code |
| `F_t` | | ✓ | | To specify the temporary label of each vertex in DHLP-1 algorithm |
| `Y_prim` | | ✓ | ✓ | To specify the y' vertex value |
| $\alpha$ | | ✓ | ✓ | One of the parameters that its value specifies the significance of the links between same type and different type vertices. |
| $\sigma$ | | ✓ | ✓ | One of the parameters of the algorithms that indicates the convergence |
| `vertex.getValue().getLastf()` | | ✓ | | To indicate the temporary label of the vertex in the previous iteration |
| `vertex.getValue().getf()` | | ✓ | ✓ | To get the label of the vertex |
| `vertex.getValue().getf_old()` | | ✓ | ✓ | To get the previous super-step label of the vertex |

As mentioned before, our proposed label propagation algorithms are designed as the distributed versions of the MINProp and Heter-LP algorithms. Our label propagation algorithms are implemented as follows: first, the vertex whose $y$ value (vertex.getValue.gety) is 1 (i.e., the first entity of the drugs) informs its adjacent nodes of its values through message passing, and label propagation operation is done. When this operation is finished, the $y$ value of that vertex becomes zero, and this process is repeated for all drug entities. In the next step, this is repeated for every entity of the diseases and targets. Finally, the interactions of the vertices with their adjacent nodes make up the output of the algorithms.

The `Compute()` method for implementing label propagation algorithms will be as follows:

```
Compute(vertex,messages)
```

```
1.  if(getSuper-step()!=0)
2.     flag:=early_checking(vertex,messages);
3.  if(getSuper-step()==0)
4.     vertex.setValue(new LPVertexValue(...)); //Set Vertex Values
5.     if(vertex.getValue.gety==1)
6.        PropagateMessage();
7.  Else
8.     DHLP-1(vertex,flag)  or DHLP-2(vertex,flag);
9.  vertex.voteToHalt();
```

In super-step==0, vertex values are firstly initialized. During this value initialization, $y$ value for the first entity of drugs becomes 1 and then, that entity sends its values to its adjacent vertices. If super-step!=0, a set of initial investigations is first done in early_checking(vertex, messages) function. If *flag=0* is the result of such investigations, DHLP-1 or DHLP-2 label propagation operation will be conducted, and if *flag=1*, the operation will not be carried out. Note that the initialized value of the flag is zero. Finally, by running the `vertex.voteToHalt()` command, each vertex is halted until it is reactivated by messages in the next super-step. The investigations and operations that each vertex does in early_checking function include:

1) The vertex whose $y$ is one saves the final label values sent by its adjacent nodes. If all adjacent nodes have conveyed their values, the vertex ends its task, changes $y$ value to zero, and informs the next vertex so that it changes its $y$ value to 1.
2) According to the received messages, each vertex checks whether or not it has the right to change its $y$ value to 1 and then inform its adjacent nodes.

3) In the last super-step, when running the algorithm is done, the vertices carry out mean operation for their mutual labels values and their adjacent nodes.

4) In each super-step, $y$ is one only for one vertex. All vertices save the ID of the vertex whose $yi$ is 1 in the current super-step. So, in the present super-step rather than the previous one, if $y$ of a different vertex is 1, the vertices will save it's ID.

5) It will be determined (based on flag value) whether label propagation should be done in the current super-step or not.

DHLP-1 label propagation function inspired by MINProp is as follows:

```
    DHLP-1 (vertex, flag)
```

```
1.  if (vertex.getValue().getCurrentf() == 0  && !flag && !vertex.getValue().IsEnd())
2.        y_prim = (1 – α)* vertex.getValue().gety();
3.        counter := 1
4.        for  (edge in vertex.getEdges())
5.           if (vertex.getId().get() % 3 != edge.getTargetVertexId().get() % 3)
6.               y_prim += α *
                       edge.getValue()*vertex.getValue().neighbours_f()[counter];
7.           counter = counter + 1;
8.      vertex.setValue(new LPVertexValue(...)); //update y_prim
9.      //determine message structure
10.     propagateMessage(vertex,messages)
11. else if(!flag && !vertex.getValue().IsEnd ())
12.        f_t =(1 – α) * vertex.getValue().gety_prim();
13.         counter := 1
14.       for (edge : vertex.getEdges())
15.          if (vertex.getId().get() % 3 == edge.getTargetVertexId().get() % 3)
16.              f_t += α * edge.getValue()
                       * vertex.getValue().neighbours_f()[counter];
17.          counter = counter + 1;
18.     vertex.setValue(new LPVertexValue(...)); //update f_t
```

```
19.        if (Math.abs(vertex.getValue().getCurrentf() -
              vertex.getValue().getLastf()) < σ
20.           vertex.setValue(new LPVertexValue(...)); //update f
21.           if (Math.abs(vertex.getValue().getf() -
                 vertex.getValue().getf_old()) < σ
22.              vertex.setValue(new LPVertexValue(...)); //enable end flag
23.        //determine message structure
24.         propagateMessage(vertex,messages)
```

In each super-step, either lines 1 to 10 or lines 11 to 24 will be run. In lines 1 to 10, $y'$ (Y_prim) value of the vertices are calculated based on $y0$ value of the vertex and its adjacent vertices. Then, its adjacent vertices are informed through a message. In lines 11 to 18, $f\_t$, or current, is firstly calculated based on $f$ value of the neighbor saved in the memory of each vertex. Next, if the difference between current and last $f$ is smaller than a threshold, $f$ will be equalized to current, and its value will be updated (lines 19 and 20). In lines 21 and 22, it is checked whether the $f\_old$ and $f$ difference is smaller than a threshold. If yes, label propagation operation is over. Finally, in lines 23 and 24, the vertex will propagate its values to its neighbors. DHLP-2 label propagation function inspired by Heter-LP and is its distributed implementation is as follows:

**DHLP-2 (vertex, flag)**

```
1.  if (!flag && !vertex.getValue().IsEnd())
2.    y_prim = (1 – α)* vertex.getValue().getf()
3.    counter := 1
4.    for (edge in vertex.getEdges())
5.       if (vertex.getId().get() % 3 != edge.getTargetVertexId().get() % 3)
6.          y_prim += α *
                edge.getValue()*vertex.getValue().neighbours_f()[counter];
7.       counter = counter + 1;
8.    f =(1 – α) * vertex.getValue().gety_prim();
9.    counter := 1
```

```
10.     for (edge in vertex.getEdges())
11.         if (vertex.getId().get() % 3 == edge.getTargetVertexId().get() % 3)
12.             f += α * edge.getValue() *vertex.getValue().neighbours_f()[counter];
13.         counter = counter + 1;
14.     vertex.setValue(new LPVertexValue(...)); //update f
15.     if (Math.abs(vertex.getValue().getf() - vertex.getValue().getf_old()) < σ
16.         vertex.setValue(new LPVertexValue(...)); //enable end flag
17.     //determine message structure
18.     propagateMessage(vertex,messages)
```

In lines 2 to 7, for each vertex, $y\_prim$ value is calculated based on $f$ values of its heterogeneous neighbors. In lines 8 to 13, $f$ value of the vertices will be calculated based on $y'$ value of the vertices and the $f$ values of its homogeneous neighbors. In line 14, $f$ value is updated for the vertex. In lines 15 and 16, if the difference between the current $f$ and the previous one is smaller than a threshold, label propagation operation will end. Otherwise, it will not end. In lines 17 and 18, the neighbors will be informed of the current status and values of the vertex.

### 3.5 Time Complexity

In DHLP-2 Algorithm, time complexity is calculated separately for every vertex in the $i$ th subnetwork as follows. Note that $i, j = 1,2,3$

$$O(t\left(1 + |V_i| + \sum_{j\%3 \neq i\%3} |V_j|\right))$$

Where $|V_j|$ is the number of vertices in the $j$ th subnetwork, and $|V_i|$ is the number of vertices in the $i$ th. $t$ is the number of iterations required for converging the calculations and depends on data structure and $\alpha$ and $\partial$ values.

In DHLP-1 Algorithm, $y'$ and ($fi\_t$ and $fi$) are calculated in separated super-steps. When $y'$ is calculated in one super-step, $fi$ and $fi\_t$ will be calculated in several subsequent super-steps until convergence is achieved. As a result, algorithm time complexity for each vertex belonging to $i$ th subnetwork will be as follows:

$$O(\,t\left(1 + \sum_{j\%3 \neq i\%3} |V_j| + t_i|V_i|\right))$$

Where $t_i$ and $t$ are the numbers of iterations required for reaching convergence in the inner and outer loops, respectively. Giraph divides the input into different partitions each of which is assigned to different workers and even various CPU cores within that worker. This way, it can enhance the speed by providing parallel running conditions. Each worker or core is responsible for running compute function for all of the vertices belonging to that partition, but running the vertices that are inside the partition is sequential, and there is no parallelism is this level. Considering the points mentioned above, the time complexity of DHLP-2 Algorithm is as follows:

$$O(\,t \sum_{i=1}^{3} \frac{|V_i|}{m}\left(1 + |V_i| + \sum_{j \neq i}|V_j|\right))$$

$$= O(\frac{t}{m} \sum_{i=1}^{3}\left(|V_i| + |V_i|^2 + \sum_{j \neq i}|V_i||V_j|\right))$$

Where $m$ is parallelism coefficient and depends on the number of workers as well as CPU cores. Therefore, the time complexity of DHLP-1 Algorithm is as follows:

$$O(\,t \sum_{i=1}^{3} \frac{|V_i|}{m}\left(1 + \sum_{j \neq i}|V_j| + t_i|V_i|\right))$$

$$= O(\frac{t}{m} \sum_{i=1}^{3}\left(|V_i| + \sum_{j \neq i}|V_i||V_j| + t_i|V_i|^2\right))$$

### 3.6 Regularization Framework and Proof of Convergence

The proposed label propagation algorithms are iterative. Hence, we need to prove that they will not run an infinite number of times, and they will be eventually converged. Also, we need to show that the answer to which the algorithms will finally converge is optimal and the best possible answer. In the existing articles, the Regularization Framework is developed in the process of proving the optimality of the algorithms (Hwang & Kuang, 2010; Shahreza et al., 2017).

In DHLP-2 Algorithm, the mathematical format of the iterative equations is as follows:

$$f_1^t(v) = (1-\alpha)y'^t_1(v) + \alpha \sum_{u\sim v} S_1(u.v) \times f_1^{t-1}(u) \quad (1)$$

Where, $f_1^t(v)$ is the label of vertex $v$ belonging to subnetwork 1 and within iteration $t$. $f_1^{t-1}(u)$ is the label of vertex $u$ belonging to subnetwork 1 and within iteration $t-1$. $S_1(u.v)$ is the similarity degree between vertices $u$ and $v$. $u$ and $v$ are neighbors in subnetwork 1. $y'^t_1(v)$ value can also be calculated as follows:

$$y'^t_1(v) = (1-\alpha)f_1^{t-1}(v) + \alpha \sum_{u\sim v} S_{1.2}(u.v)f_2^{t-1}(u) + \sum_{u\sim v} S_{1.3}(u.v)f_3^{t-1}(u) \quad (2)$$

In the above equation, $S_{1.2}(u.v)$ represents the similarity degree between $u$ and $v$ neighboring vertices where $u$ belongs to subnetwork 1 and $v$ belongs to subnetwork 2. In addition, $S_{1.3}(u.v)$ indicates the similarity degree between $u$ and $v$ vertices where $u$ belongs to subnetwork 1 and $v$ belongs to subnetwork 3.

Considering equations 1 and 2 and $(1-\alpha) = \beta$, we have:

$$f_1^t(v) = \beta(\beta y_1(v) + \alpha S_{1.2}f_2^{t-1} + \alpha S_{1.3}f_3^{t-1}) + \alpha S_1 f_1(t-1) \quad (3)$$

Where $S_{1.2}$ is a $1 \times n2$ vector representing the degree of similarity between vertex $v$ and $n2$ of its vertices in subnetwork 2. Also, $S_{1.3}$ is a $1 \times n3$ vector indicating the degree of similarity between vertex $v$ and $n3$ of its vertices in subnetwork 3. $f2$ and $f3$ are also vectors with $n2 \times 1$ and $n3 \times 1$ dimensions containing vertices labels of subnetworks 2 and 3.

The third equation is equivalent to the iterative equation of (Shahreza, Ghadiri, Mousavi, Varshosaz, & Green, 2017). So, the proof of convergence, as well as optimality of this algorithm, will be equivalent to (Shahreza, Ghadiri, Mousavi, Varshosaz, & Green, 2017). Likewise, for algorithm DHLP-1, it will be determined that its iterative equations are similar to iterative equations of (Hwang & Kuang, 2010). So, the proof of convergence and optimality will be the same.

## 4 Results

We designed a comprehensive set of experiments to evaluate the performance of the proposed DHLP-1 and DHLP2- algorithms. The primary criteria are speed and accuracy that are evaluated a real-world application domain in bioinformatics called "drug repositioning". In drug repositioning, such algorithms help the domain experts to find new usage for already-approved drugs.

## 4.1 Computing Environment

The experiments were conducted by Hadoop Cluster consisting of nine computer nodes. One of the nodes was the master, while the others were slaves. The specifications of the master node were Intel Core i5-4460 CPU (3.2GHZ), 8 GB of RAM, 1 TB of hard disk space, CentOS 7 operating system, and each slave node included Intel Core i7-2600 CPU (3.4 GHz), 4 GB of RAM, 1 TB of hard disk space, CentOS 7. All configurations took advantage of Hadoop 2.7.1 and Apache Giraph 1.3.0. The nodes were connected through a 100Mbps local area network. Considering the architectural nature of Giraph that is based on worker-slave, some of the nodes were required to be selected as masters (whether active or dominant) and some as workers. Therefore, two of the nodes were masters and six of them were workers in the experiments.

## 4.2 Statistical Analysis

We used three metrics to evaluate the prediction accuracy of interactions. These measures have been widely used in drug repositioning studies.

1) Area Under the Curve (AUC) of the Receiver Operating Characteristics (ROC) Curve: ROC Curve indicates the number of TPs in proportion to the number of FPs based on various values of decision threshold. TP is the interactions indeed predicted, while FP is the false positive interactions, i.e., there have been no predicted interactions in the real world. AUC is the area under the ROC Curve.
2) Area Under the Precision-Recall (AUPR) Curve: Precision is the proportion of truly predicted points to the number of points ranked above a certain limit. Recall is the proportion of truly predicted points to the total number of points that are true in the real world. AUPR is a parameter representing the proportion under the Precision-Recall Curve.
3) Best Accuracy: Accuracy measures the difference between the estimated and real value. It is determined as follows:

$$\text{Accuracy} = \frac{TP+TN}{TP+FP+TN+FN}$$

Where FN=False Negative, FP=False Positive, TN=True Negative, and TP=True Positive. Best Accuracy is the highest value of accuracy obtained while repeating the experiment for several times with the same parameters and data. We use Best Accuracy to show how well our algorithms perform in best and ideal cases.

### 4.2.1 Evaluation of Accuracy Based on 10-Fold Cross-Validation

We used 10-fold cross-validation to analyze the accuracy of the proposed algorithms and compare with the non-distributed versions. So, the original data were divided into ten parts. Nine parts are used for training and one part for testing. Table 2 represents the results after running DHLP-1, DHLP-2, Heter-LP, and MinProp algorithms on GPCR data which is a group of four protein targets besides Enzyme, Ion Channel, and Nuclear Receptor (Yamanishi, Araki, Gutteridge, Honda, & Kanehisa, 2008). In general, DHLP-1 and DHLP-2 show better accuracy than MINProp and Heter-LP, and the reason is that in both centralized and distributed algorithms, $\sigma$ is a threshold that controls the accuracy of the results and determines the number of iterations required to reach convergence. In the centralized algorithms, the checking of $\sigma$ is vector-based for the group of entities in one subnetwork. However, in the distributed versions, it is node-based per individual in a subnetwork. In other words, distributed versions ensure accuracy for every single individual of entities which means, it can provide higher overall accuracy. Furthermore, distributed algorithms are iterative, and while the accuracy threshold for at least one of the entities is not satisfied, it continues to do its iterative processing. Hence, it runs more iterations. More iterations ensure a better result in terms of accuracy.

*Table 2. Average results of 10-fold CV for AUC, AUPR, and BestACC in four algorithms of DHLP-1, DHLP-2, Heter-LP, and MINProp on GPCR data, $\alpha$= 0.5, $\sigma$ = 0.5*

| Interaction | Algorithm | AUC | AUPR | BestACC |
|---|---|---|---|---|
| Drug-disease | DHLP-1 | 0.9520 | 0.9687 | 0.9390 |
| | DHLP-2 | **0.9527** | **0.9760** | **0.9531** |
| | MINProp | 0.5 | 0.2478 | 0.5042 |
| | HeterLP | 0.79321 | 0.83176 | 0.72315 |
| Drug-target | DHLP-1 | **0.9757** | 0.7661 | 0.9848 |
| | DHLP-2 | 0.9549 | **0.9555** | **0.9972** |
| | MINProp | 0.5 | 0.0149 | 0.97 |
| | HeterLP | 0.96747 | 0.79579 | 0.98584 |
| Disease-target | DHLP-1 | 0.9510 | 0.9691 | 0.9529 |
| | DHLP-2 | **0.9530** | **0.9762** | **0.9531** |
| | MINProp | 0.5 | 0.2491 | 0.5016 |
| | HeterLP | 0.79582 | 0.69718 | 0.84017 |

### 4.2.2 Prediction of a Deleted Interaction

A well-known approach for validating the algorithms is to remove some of the interactions from the input data set, then to run the process to recover the deleted interactions (as one does during a cross-validation experiment). This approach will determine whether those algorithms can predict deleted interactions or not. To this aim, we conducted two distinct experiments using the drug-target subnetwork. In the first experiment, we chose an arbitrary drug and removed one of its interactions from the input data set, and in the second one, we excluded all of its interactions. The experiments have been carried out for both DHLP-1 and DHLP-2 algorithms. Many tests have been done on this issue, one of which is discussed here as a case study. The ability of algorithms in predicting new interactions was evaluated in the first experiment. D00232 is a drug with three interactions with targets hsa: 1128, hsa:1129, and hsa: 1131 targets where they are target IDs with the official full name of cholinergic receptor muscarinic 1, cholinergic receptor muscarinic 2 cholinergic receptor muscarinic 3 respectively. We deleted its interaction with hsa: 1128 from the input, and then executed DHLP-1 and DHLP-2. As shown in the results, both algorithms have been able to predict the deleted relation correctly. Table 3 represents 20 new top-ranked interactions predicted by both algorithms.

*Table 3. Investigating the removal of hsa:1128 interaction for D00232 drug and the fact that whether or not the algorithms have the ability to predict the removed interaction, and 20 predicted top-rank relations for this drug after running DHLP-1 and DHLP-2 algorithms, α= 0.5, σ = 0.5*

| NO. | DHLP-1 | DHLP-2 |
|---|---|---|
| 1 | **hsa:1129** | **hsa:1129** |
| 2 | **hsa:1131** | **hsa:1128** |
| 3 | **hsa:1128** | **hsa:1131** |
| 4 | hsa:3269 | hsa:154 |
| 5 | hsa:154 | hsa:3269 |
| 6 | hsa:153 | hsa:153 |
| 7 | hsa:1813 | hsa:1813 |
| 8 | hsa:4988 | hsa:4988 |
| 9 | hsa:148 | hsa:148 |
| 10 | hsa:1132 | hsa:1132 |
| 11 | hsa:1133 | hsa:1133 |
| 12 | hsa:150 | hsa:185 |
| 13 | hsa:185 | hsa:3274 |
| 14 | hsa:3274 | hsa:150 |

| 15 | hsa:3577 | hsa:3577 |
| 16 | hsa:1814 | hsa:1814 |
| 17 | hsa:3356 | hsa:3360 |
| 18 | hsa:155 | hsa:146 |
| 19 | hsa:146 | hsa:5737 |
| 20 | hsa:147 | hsa:7201 |

### 4.2.3 Prediction of Pseudo New Drugs

In the second experiment, we generated a new drug by removing all of the interactions of it with its targets. This experiment aims to investigate the abilities of the algorithms in predicting the interactions of a new drug.

D00232 has three interactions with targets in the input data set. In the experiment, all three interactions of D00232 were deleted, and DHLP-1 and DHLP-2 algorithms were executed. Table 4 indicates the interactions discovered after the execution. As can be seen, both algorithms were able to predict the three removed interactions.

*Table 4. Investigating the removal of all interactions for D00232 drug and the fact that whether or not the algorithms have the ability to predict the removed interactions, and 20 predicted top-rank relations for this drug after running DHLP-1 and DHLP-2 algorithms, α= 0.5, σ = 0.5*

| NO. | *DHLP-1* | *DHLP-2* |
|---|---|---|
| 1 | **hsa:1128** | **hsa:1128** |
| 2 | hsa:154 | hsa:154 |
| 3 | hsa:3269 | hsa:3269 |
| 4 | hsa:153 | hsa:153 |
| 5 | hsa:1813 | hsa:148 |
| 6 | **hsa:1129** | hsa:1813 |
| 7 | hsa:4988 | **hsa:1129** |
| 8 | hsa:148 | hsa:4988 |
| 9 | hsa:185 | hsa:185 |
| 10 | hsa:150 | hsa:150 |
| 11 | hsa:3274 | hsa:3577 |
| 12 | hsa:3577 | hsa:3274 |
| 13 | hsa:1814 | hsa:1814 |
| 14 | **hsa:1131** | hsa:146 |

| 15 | hsa:5737 | hsa:5739 |
|---|---|---|
| 16 | hsa:155 | hsa:5737 |
| 17 | hsa:3356 | hsa:3360 |
| 18 | hsa:146 | hsa:5732 |
| 19 | hsa:147 | **hsa:1131** |
| 20 | hsa:3360 | hsa:5731 |

## 4.3 Performance Analysis

In this section, the efficiency of the algorithms regarding speed is evaluated. In this regard, four experiments were designed for: 1) The effect of multithreading, 2) the effect of the number of workers, 3) the speed obtained in distributed heterogeneous label propagation algorithms compared to non-distributed versions of them and 4) the effect of the value of σ were investigated.

### 4.3.1 The effect of multithreading on runtime

We used multithreading embedded in Giraph to improve the runtime and scalability of the algorithms. Figure 3 represents the effect of an increase in the number of threads on the runtime of algorithms. As the number of threads has increased to 8 in algorithm DHLP-1, the runtime has decreased and then has remained unchanged. In algorithm DHLP-2, as the number of threads has increased to four, the running time has decreased and then has remained unchanged. Since each slave node has Core i7-2600 CPU, it only supports 8 hyper threads and at maximum regardless of the Giraph's ability to employ as many threads as we want, the number of effective threads will be 8 and for more than 8 threads, the results would be same as 8 threads.

### 4.3.2 The effect of the number of workers on runtime

In Figure 4, the effect of increasing the number of workers (from one to six) on the runtime of the algorithm is measured. In both algorithms, as the number of workers has increased, the runtime has decreased. However, the slope of decrease in execution time for DHLP-2 was lower. The significant running time scale difference for figures 3 and 4 comes from differences between figures 3 and 4's set up in terms of dataset type and size and the number of running super-steps. In figure 3, the experiment was conducted for only 200 super-steps or in other word a partial running but in figure 4, the experiment was conducted for complete running.

### 4.3.3 Comparing distributed and non-distributed label propagation algorithms runtime

DHLP-1 and DHLP-2 algorithms are the distributed versions of MinProp and Heter-LP, respectively. Comparison of the speed between the distributed and non-distributed versions according to the different number of edges (from 1M to 20M) of the heterogeneous network has been presented in Tables 5 and 6. Since the number of workers is 6 and the number of threads is 8, the maximum theoretically possible speed up without considering the networking delay is 48. In our experiments we could achieve a maximum speed up of 4.8 for DHLP-1 and a maximum speed up of 11.22 for DHLP-2 both for 20M edges. Note that experiment setup for non-distributed versions (MinProp and Heter-LP) is same as slave nodes which is Intel Core i7-2600 CPU (3.4 GHz), 4 GB of RAM, 1 TB of hard disk space, CentOS 7.

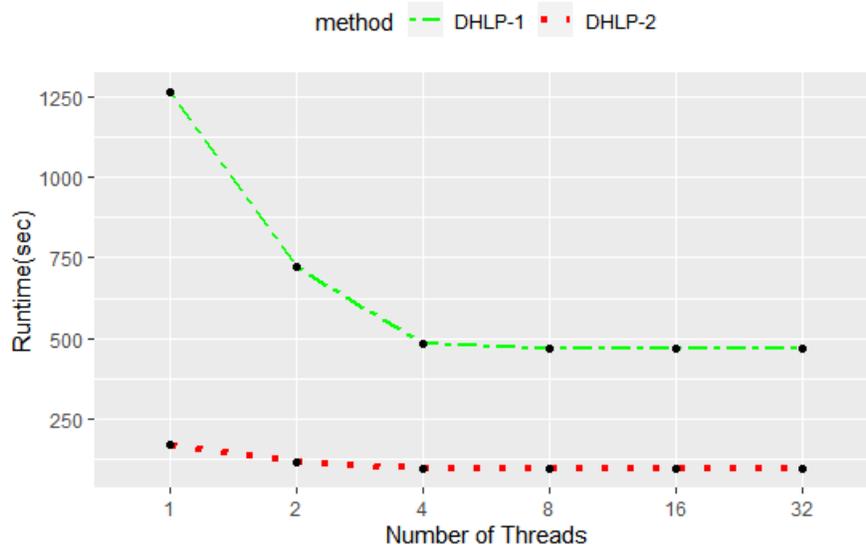

*Figure 3. Investigating the effect of the number of threads for Algorithms - Running Time for 200 Super-steps – 20 milion edge - 1 node – α= 0.5, σ = 0.5*

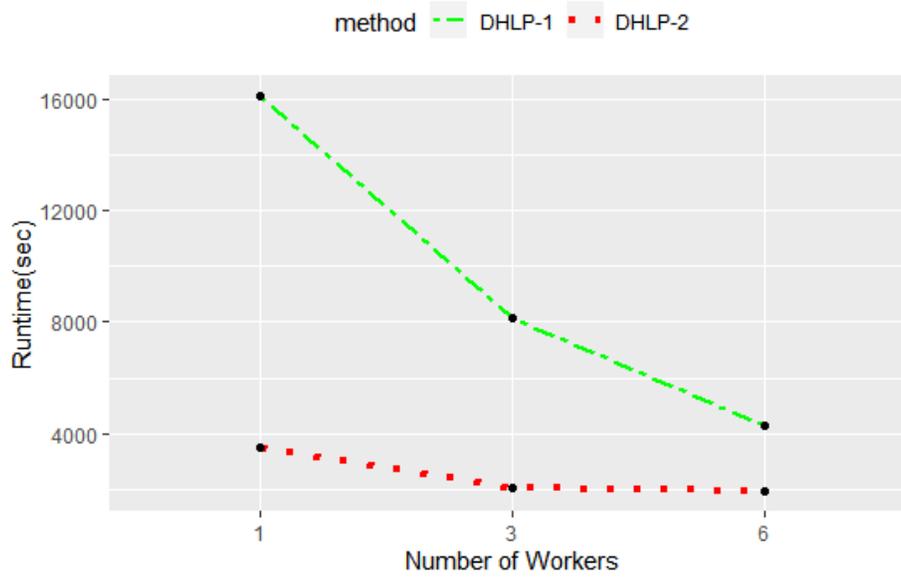

*Figure 4. Investigating the effect of the number of workers on algorithms running time for 1 thread; 10 million edges for DHLP-2; 5 million edges for DHLP-1, α= 0.5, σ = 0.5*

*Table 5. Gain obtained with DHLP-1 for different datasets, α= 0.5, σ = 0.5, 6 workers, 8 threads*

| Number of Edges | Minprop (s) | DHLP-1 (s) | Gain (Minprop/DHLP-1) |
|---|---|---|---|
| 1M | 664 | 323 | **2.06** |
| 2M | 1852 | 635 | **2.92** |
| 3M | 3591 | 1028 | **3.49** |
| 4M | 5704 | 1503 | **3.8** |
| 5M | 7816 | 2039 | **3.83** |
| 6M | 10649 | 2670 | **3.99** |
| 7M | 13535 | 3232 | **4.19** |
| 8M | 16066 | 3966 | **4.05** |
| 9M | 19761 | 4464 | **4.43** |
| 10M | 23342 | 5161 | **4.52** |
| 20M | 66719 | 13894 | **4.80** |

*Table 6. Gain obtained with DHLP-2 for different datasets, α= 0.5, σ = 0.5, 6 worker, 8 threads*

| Number of Edges | HeterLP (s) | DHLP-2 (s) | Gain (HeterLP/DHLP-2) |
|---|---|---|---|
| 1M | 700 | 153 | **4.58** |
| 2M | 2015 | 334 | **6.03** |
| 3M | 3768 | 545 | **6.91** |
| 4M | 5633 | 726 | **7.76** |
| 5M | 7910 | 873 | **9.06** |
| 6M | 10844 | 1106 | **9.80** |
| 7M | 13654 | 1224 | **11.15** |
| 8M | 16351 | 1570 | **10.41** |
| 9M | 20246 | 2036 | **9.94** |
| 10M | 23976 | 2289 | **10.47** |
| 20M | 68899 | 6140 | **11.22** |

### 4.3.4 The effect of σ on runtime

One of the useful parameters in the algorithms is $\sigma$. The effect of the decreasing this value in increase of the runtime is presented in Table 7.

*Table 7. Investigating the effect of σ on convergence rate for small data (GPCR) with 6 nodes and 8 threads*

| Dataset | Algorithm | Parameters | $\sigma = 0.2$ (s) | $\sigma = 0.1$ (s) | $\sigma = 0.05$ (s) | $\sigma = 0.01$ (s) | $\sigma = 0.005$ (s) | $\sigma = 0.002$ (s) |
|---|---|---|---|---|---|---|---|---|
| GPCR | DHLP-1 | 8 Thread 6 worker $\alpha = 0.5$ | 120 | 127 | 133 | 151 | 159 | 223 |
| | DHLP-2 | 8 Thread 6 worker $\alpha = 0.5$ | 69 | 75 | 82 | 94 | 102 | 125 |

## 5 Discussion

In this paper, with utilizing concepts of label propagation and distributed computing, DHLP-1 and DHLP-2 are presented for distributed label propagation in heterogeneous networks. As a case study, these two algorithms are employed in drug repositioning which is an important field in bioinformatics and biology area in order to evaluate their effectiveness. The experiments are designed in two parts 1) Statistical Analysis 2) Performance Analysis.

In statistical analysis, evaluation of Accuracy is done based on 10-Fold Cross-Validation and is shown in Table 2. According to the table, DHLP-1 and DHLP-2 are more accurate than MinProp and Heter-LP in most cases, and except in one case, the obtained accuracy of DHLP-2 is more than that of DHLP-1. MinProp has the worst performance and lowest accuracy among others. Moreover, based on experiments shown in Tables 3 and 4, proposed algorithms have the ability to predict the new drugs, diseases and targets interactions. They have also the ability to predict removed interactions.

In performance analysis, the effect of multithreading on runtime is investigated. As shown in Figure 3 due to an overhead increase with the increase in the number of threads, after the certain number of threads the runtime remained unchanged in DHP-1 and has increased in DHLP-2. In both algorithms, the slope of decrease in time is high at first and low afterward due to the overload increase.

In Figure 4, the effect of the number of workers on runtime is measured. In both algorithms, increase in the number of workers leads to decrease in runtime but the slope of decreasing for DHLP-2 is lower because the DHLP-2 algorithm is more straightforward and not time-consuming as DHLP-1. Thus, as the number of workers increases, the intensity of the decrease is less because the communication cost between more workers will be higher.

In Tables 5 and 6, comparison of the runtime between distributed and non-distributed label propagation algorithms is presented. Gain is the runtime proportion of distributed version to non-distributed version. Experiments is revealing that the obtained Gain for DHLP-2 is higher than the one for DHLP-1. Furthermore, this proportion grows as the number of edges grows. That is, in larger networks, the difference will be bigger and DHLP-1 and DHLP-2 algorithms will be more efficient due to initialization overhead for smaller networks. Moreover, the distributed is being carried out on 6 workers. As the number of workers grows, the obtained gain is supposed to increase as well. An important observation from the results is that the runtime increase is not linear with the growing of number of edges and the reason is that the part of total running time consists of networking time and communication between computer nodes and networking time is increased as well with the growing of number of edges. In addition, according to the results of Tables 5 and

6, our distributed algorithms offer scalable solutions which perform comparatively better against non-distributed versions in larger networks.

The effect of $\sigma$ on runtime is investigated in Table 7. As $\sigma$ decreases, the runtime increases. This occurs because in smaller $\sigma$, convergence happens late. Therefore, runtime is more. Another parameter is $\alpha$ whose increase or decrease does not cause any special pattern in the runtime, for it is related to data structure and shows different behavior in different data.

In general, our investigation suggests the high effectiveness of the proposed algorithms. As a future work, the usability of the algorithms might be examined in domains other than drug repositioning or applied to other heterogeneous networks. Moreover, providing a bigger cluster, the runtime could be lower, and such algorithms may be used for larger networks.

# 6 Conclusion

In this paper, we introduced two distributed algorithms, namely DHLP-1 and DHLP-2, for label propagation in heterogeneous networks. The algorithms were theoretically explained, mathematically proved, and exploited in drug repositioning domain as a case study so that their effectiveness is evaluated. Based on the comparison between the distributed and non-distributed versions, the distributed versions of the algorithms lead to great scalability and decrease in the runtime. Experiments of effectiveness analysis were conducted using 10-fold cross-validation test. The obtained AUC, AUPR, and BestAccuracy indicate the high accuracy of the proposed algorithms that are significantly better than their non-distributed versions. In addition, empirical experiments demonstrated that the proposed algorithms performed well in predicting new drugs, targets, and diseases interactions.

**References**

Alaimo, S., Pulvirenti, A., Giugno, R., & Ferro, A. (2013). Drug–target interaction prediction through domain-tuned network-based inference. *Bioinformatics, 29*(16), 2004-2008.
Andronis, C., Sharma, A., Virvilis, V., Deftereos, S., & Persidis, A. (2011). Literature mining, ontologies and information visualization for drug repurposing. *Briefings in bioinformatics, 12*(4), 357-368.
Bhat, A. U. (2012). Scalable community detection using label propagation & map-reduce. *Scalable Community Detection using Label Propagation\& Map-Reduce*.
Bhuiyan, H., Khan, M., Chen, J., & Marathe, M. (2017). Parallel algorithms for switching edges in heterogeneous graphs. *Journal of Parallel and Distributed Computing, 104*, 19-35.
Chen, B., Ding, Y., & Wild, D. J. (2012). Assessing drug target association using semantic linked data. *PLoS computational biology, 8*(7), e1002574.
Chen, H., & Zhang, Z. (2013). A semi-supervised method for drug-target interaction prediction with consistency in networks. *PLoS one, 8*(5), e62975.

Chen, X., Liu, M.-X., & Yan, G.-Y. (2012). Drug–target interaction prediction by random walk on the heterogeneous network. *Molecular BioSystems, 8*(7), 1970-1978.
Ching, A., Edunov, S., Kabiljo, M., Logothetis, D., & Muthukrishnan, S. (2015). One trillion edges: Graph processing at facebook-scale. *Proceedings of the VLDB Endowment, 8*(12), 1804-1815.
Farhangi, E., Ghadiri, N., Asadi, M., Nikbakht, M. A., & Pitre, S. (2017). *Fast and scalable protein motif sequence clustering based on Hadoop framework.* Paper presented at the 2017 3th International Conference on Web Research (ICWR).
Gottlieb, A., Stein, G. Y., Ruppin, E., & Sharan, R. (2011). PREDICT: a method for inferring novel drug indications with application to personalized medicine. *Molecular systems biology, 7*(1).
Grady, L. (2006). Random walks for image segmentation. *IEEE Transactions on Pattern Analysis & Machine Intelligence*(11), 1768-1783.
Gregory, S. (2010). Finding overlapping communities in networks by label propagation. *New Journal of Physics, 12*(10), 103018.
Hwang, T., & Kuang, R. (2010). *A heterogeneous label propagation algorithm for disease gene discovery.* Paper presented at the Proceedings of the 2010 SIAM International Conference on Data Mining.
Kajdanowicz, T., Kazienko, P., & Indyk, W. (2014). Parallel processing of large graphs. *Future Generation Computer Systems, 32*, 324-337.
Li, J., & Lu, Z. (2012). *A new method for computational drug repositioning using drug pairwise similarity.* Paper presented at the 2012 IEEE International Conference on Bioinformatics and Biomedicine.
Liu, J., Xu, B., Xu, X., & Xin, T. (2016). A link prediction algorithm based on label propagation. *Journal of computational science, 16*, 43-50.
Maleki, E. F., Azadani, M. N., & Ghadiri, N. (2016). *Performance evaluation of spatialhadoop for big web mapping data.* Paper presented at the 2016 Second International Conference on Web Research (ICWR).
Malewicz, G., Austern, M. H., Bik, A. J., Dehnert, J. C., Horn, I., Leiser, N., & Czajkowski, G. (2010). *Pregel: a system for large-scale graph processing.* Paper presented at the Proceedings of the 2010 ACM SIGMOD International Conference on Management of data.
Martella, C., & Shaposhnik, R. *Practical graph analytics with apache giraph* (Vol. 1): Springer.
Menden, M. P., Iorio, F., Garnett, M., McDermott, U., Benes, C. H., Ballester, P. J., & Saez-Rodriguez, J. (2013). Machine learning prediction of cancer cell sensitivity to drugs based on genomic and chemical properties. *PLoS one, 8*(4), e61318.
Mohan, A., Venkatesan, R., & Pramod, K. (2017). A scalable method for link prediction in large real world networks. *Journal of Parallel and Distributed Computing, 109*, 89-101.
Napolitano, F., Zhao, Y., Moreira, V. M., Tagliaferri, R., Kere, J., D'Amato, M., & Greco, D. (2013). Drug repositioning: a machine-learning approach through data integration. *Journal of cheminformatics, 5*(1), 30.
Rossi, R. G., de Andrade Lopes, A., & Rezende, S. O. (2016). Optimization and label propagation in bipartite heterogeneous networks to improve transductive classification of texts. *Information Processing & Management, 52*(2), 217-257.
Rossi, R. G., Lopes, A. A., & Rezende, S. O. (2014). *A parameter-free label propagation algorithm using bipartite heterogeneous networks for text classification.* Paper presented at the Proceedings of the 29th Annual ACM Symposium on Applied Computing.
Shahreza, M. L., Ghadiri, N., Mousavi, S. R., Varshosaz, J., & Green, J. R. (2017). Heter-LP: A heterogeneous label propagation algorithm and its application in drug repositioning. *Journal of biomedical informatics, 68*, 167-183.
Shi, C., Li, Y., Zhang, J., Sun, Y., & Philip, S. Y. (2016). A survey of heterogeneous information network analysis. *IEEE Transactions on Knowledge and Data Engineering, 29*(1), 17-37.
Silva, T. C., & Zhao, L. (2016). *Machine learning in complex networks* (Vol. 2016): Springer.
Stanisz, T., Kwapień, J., & Drożdż, S. (2019). Linguistic data mining with complex networks: A stylometric-oriented approach. *Information Sciences, 482*, 301-320.

Tari, L. B., & Patel, J. H. (2014). Systematic drug repurposing through text mining. In *Biomedical Literature Mining* (pp. 253-267): Springer.
Tian, Z., & Kuang, R. (2012). *Global linear neighborhoods for efficient label propagation.* Paper presented at the Proceedings of the 2012 SIAM International Conference on Data Mining.
Valiant, L. G. (1990). A bridging model for parallel computation. *Communications of the ACM, 33*(8), 103-111.
Wang, W., Yang, S., & Li, J. (2013). Drug target predictions based on heterogeneous graph inference. In *Biocomputing 2013* (pp. 53-64): World Scientific.

Xia, Z., Wu, L.-Y., Zhou, X., & Wong, S. T. (2010). *Semi-supervised drug-protein interaction prediction from heterogeneous biological spaces.* Paper presented at the BMC systems biology.

Xie, J., & Szymanski, B. K. (2013). *Labelrank: A stabilized label propagation algorithm for community detection in networks.* Paper presented at the 2013 IEEE 2nd Network Science Workshop (NSW).

Yamanishi, Y., Araki, M., Gutteridge, A., Honda, W., & Kanehisa, M. (2008). Prediction of drug–target interaction networks from the integration of chemical and genomic spaces. *Bioinformatics, 24*(13), i232-i240.

Yan, X.-Y., Zhang, S.-W., & Zhang, S.-Y. (2016). Prediction of drug–target interaction by label propagation with mutual interaction information derived from heterogeneous network. *Molecular BioSystems, 12*(2), 520-531.

Zhang, P., Wang, F., & Hu, J. (2014). *Towards drug repositioning: a unified computational framework for integrating multiple aspects of drug similarity and disease similarity.* Paper presented at the AMIA Annual Symposium Proceedings.

Zhou, D., Bousquet, O., Lal, T. N., Weston, J., & Schölkopf, B. (2004). *Learning with local and global consistency.* Paper presented at the Advances in neural information processing systems.

Zhu, Q., Tao, C., Shen, F., & Chute, C. G. (2014). Exploring the pharmacogenomics knowledge base (PharmGKB) for repositioning breast cancer drugs by leveraging Web ontology language (OWL) and cheminformatics approaches. In *Biocomputing 2014* (pp. 172-182): World Scientific.